\documentclass[twocolumn]{aa} 
\usepackage{graphicx}
\usepackage{txfonts}
\usepackage{natbib}
\usepackage[dvips]{color}
\bibpunct{(}{)}{,}{a}{}{,}

\def\ha{{H$\alpha$}\/} 
\def\lya{{Ly$\alpha$}\/}
\def\hb{{H$\beta$}\/} 
 
\def\Oii{{[{{\sc O}\,{\sc ii}}]~$\lambda\lambda$3726,3728}\/} 
\def\Oiii{{[{{\sc O}\,{\sc iii}}]~$\lambda\lambda$4959,5007}\/}

\def\NII{{[{{\sc N}\,{\sc ii}}]}\/}
\def\HI{{\sc Hi}\/}
\def\HII{{\sc Hii}\/}
\def\na{{$N_{810}$}\/}
\def\nb{{$N_{817}$}\/}
\def\nc{{$N_{824}$}\/}
\def\med{{$M_{815}$}\/}

\def\simlt{\lower.5ex\hbox{$\; \buildrel < \over \sim \;$}}
\def\simgt{\lower.5ex\hbox{$\; \buildrel > \over \sim \;$}}

\def\taurus2{{\sc Taurus-2}}
\def\lineunits{erg\,s$^{-1}$\,cm$^{-2}$}
\def\ergsscm{erg\,s$^{-1}$\,cm$^{-2}$}

\def\ergsPerAng{erg\,s$^{-1}$\,cm$^{-2}$\,\AA$^{-1}$}
\def\ergs2band{erg\,s$^{-1}$\,cm$^{-2}$\,band$^{-1}$}
\def\ergsPersec{erg\,s$^{-1}$}

\def\am{$'$}

\def\MpcPer3{Mpc$^{-3}$}
\def\Mpc3{Mpc$^{3}$}

\def\kmsMpc{km\,s$^{-1}$\,Mpc$^{-1}$}

\def\sqdeg{\,$\Box^\circ$}

\def\kms{km\,s$^{-1}$}

\def\Msunyr{M$_\odot$\,yr$^{-1}$}

\begin{document}
\title{The Wide Field Imager Lyman-Alpha Search (WFILAS) for Galaxies
  at Redshift $\sim$\,5.7 \thanks{Based on observations taken at the
    Cerro La Silla (ESO programs 67.A-0063, 68.A-0363, 69.A-0314 and
    MPG time) and Cerro Paranal Observatories (ESO program
    272.A-5029).}  }

\subtitle{I. A Spatially Compact \lya\ Emitting Galaxy at Redshift
  5.721}

\author{Eduard Westra\inst{1}, D.\ Heath Jones\inst{1}, Chris E.\ 
  Lidman\inst{2}, Ramana M. Athreya\inst{3}, Klaus
  Meisenheimer\inst{4}, Christian Wolf\inst{5}, Thomas
  Szeifert\inst{2}, Emanuela Pompei\inst{2} \and Leonardo
  Vanzi\inst{2} }

\offprints{Eduard Westra, e-mail: {\tt westra@mso.anu.edu.au}}

\institute{Research School of Astronomy \& Astrophysics, The
  Australian National University, Weston Creek ACT 2611, Australia,
  \email{westra@mso.anu.edu.au, heath@mso.anu.edu.au}
  \and European Southern Observatory, Casilla 19001, Santiago 19,
  Chile, \email{clidman@eso.org, tszeifer@eso.org, epompei@eso.org,
    lvanzi@eso.org}
  \and {National Centre for Radio Astrophysics, Tata Institute of
    Fundamental Research Pune University Campus, Post Bag 3,
    Ganeshkhind Pune 411007, India, \email{rathreya@ncra.tifr.res.in}}
  \and {Max Planck Institute f\"ur Astronomie, K\"onigstuhl 17,
    D-69117 Heidelberg, Germany, \email{meise@mpia.de}}
  \and {Department of Astrophysics, Denys Wilkinson Building,
    University of Oxford, Keble Road, Oxford, OX1 3RH, U.K.,
    \email{cwolf@astro.ox.ac.uk}}
}

\date{Received ---; accepted ---}

\abstract{We report the spectroscopic confirmation of a compact \lya\ 
  emitting galaxy at $z=5.721$. A FORS2 spectrum of the source shows a
  strong asymmetric line with a flux of 5$\times$10$^{-17}$
  erg\,s$^{-1}$\,cm$^{-2}$, making it one of the brightest \lya\ 
  emitting galaxies at this redshift, and a line-of-sight velocity
  dispersion of 400\,km\,s$^{-1}$. We also have a tentative detection
  of a second, narrower component that is redshifted by
  400\,km\,s$^{-1}$ with respect to the main peak. A FORS2 image shows
  that the source is compact, with a FWHM of 0\farcs5, which
  corresponds to 3.2 kpc at this redshift\thanks{Throughout this
    Letter, a cosmology with $H_0$ = 70~\kmsMpc, $\Omega_{\rm M}$ =
    0.3 and $\Omega_\Lambda$ = 0.7 is assumed. Magnitudes are on the
    AB-system.}. This source is a brighter example of
  \object{J1236.8+6215} \citep{Dawson02}, another \lya\ emitting
  galaxy at $z\sim5.2$.
  
  }

\authorrunning{Westra et al.}

\titlerunning{A compact \lya\ emitting galaxy at $z \sim$\,5.7}

\maketitle

\section{Introduction}

Wide-field imaging surveys with specially selected narrow-band filters
are an effective means of discovering high redshift ($z \gtrsim 5$)
\lya\ emitting galaxies \citep[see][ and references therein]{Hu04}.
Spectra of these galaxies are dominated by a single, asymmetric
emission line. One of the strongest arguments for associating this
line with \mbox{Lyman-$\alpha$} (\lya) is the asymmetry in the line
profile \citep[e.g.][]{Stern00}, which can only be detected if the
spectral resolution is high enough ($R \gtrsim 2000$).

The profile of the \lya\ line is the end result of emission from \HII\ 
regions and resonant scattering by \HI. The bulk of the \lya\ emission
comes from the recombination of hydrogen that has been ionised by UV
flux of massive stars. Part of the ionisation may be due to shocks
\citep{BlandHawthorn04} or from an AGN, although in a study of \lya\ 
emitting galaxies at $z\sim4.5$, \citet{Wang04} found no evidence of
AGN activity. The shape of the line profile is sensitive to the
geometry, density and kinematics of both the \HII\ gas, where the line
is produced, and the \HI\ gas, where it is scattered
\citep{Ahn03,Santos04}. Dust can also play a role. Hence, the \lya\ 
emission line can be viewed as a tool, albeit a rather blunt one, that
might be used to constrain the spatial and kinematic distribution of
the hydrogen gas in these distant galaxies.

In this letter, we present the 0\farcs5 resolution seeing-limited
imaging and $R \sim 3600$ spectroscopy of a \lya\ emitting galaxy that
was selected from \lya\ emitting candidates in the WFILAS catalog
(Westra et al. 2005, in prep.).

\section{WFILAS and Candidate Selection}
\label{sec:observations}

\begin{figure*}[!htbp]
  \centering
  \includegraphics[width=\textwidth, trim=3pt 2pt 153pt 686pt, clip]{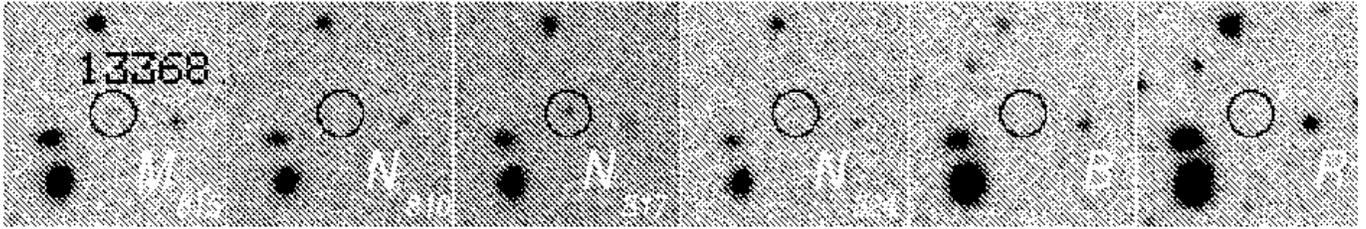}
  \caption{
    From left to right, thumbnails of the confirmed \lya\ emitter
    \object{S11\_13368} at redshift 5.721. Each thumbnail covers a
    24\arcsec\ $\times$ 24\arcsec\ region with a pixel scale of
    0\farcs238\,pix$^{-1}$. The emitter is only detected in the \med\ 
    and \nb\ images.}
  \label{fig:thumbs}
\end{figure*}

WFILAS is a survey for bright \lya\ emitting galaxies at $z \sim 5.7$.
The selection strategy is similar to the successful strategies
employed in other \lya\ surveys at these redshifts
\citep{Ajiki03,Rhoads03,Hu04}. WFILAS covers a larger volume
($\sim$1.2$\times$10$^{6}$\,\Mpc3), a larger area ($\sim$1\sqdeg) and
has a brighter detection limit (2$\sigma$ limit magnitude
$\sim$24.0--24.5) than these surveys. Hence the candidate \lya\ 
emitting galaxies in the WFILAS catalogue will, on average, be
brighter than the \lya\ emitting galaxies in these other catalogs.

The survey used the Wide Field Imager (WFI) on the ESO/MPI 2.2m
telescope at the Cerro La Silla Observatory and targeted three fields.
The WFI consists of a mosaic of eight \mbox{(4 x 2)} \mbox{2k x 4k}
CCDs arranged to give a field of view of \mbox{34\am\ x 33\am\ }with a
pixel scale of 0\farcs238 pixel$^{-1}$. Images were taken with the
standard broad-band $B$ and $R$ filters, and four narrower filters -
an intermediate-band ($\Delta\lambda$ = 22\,\AA) filter at 815\,nm
(\med) and three custom-made narrow-band ($\Delta\lambda$ = 7\,\AA)
filters with central wavelengths at 810\,nm (\na), 817\,nm (\nb) and
824\,nm (\nc).  They lie in a spectral region where the emissivity of
the night-sky is relatively low, which improves the sensitivity to
\lya\ emission.

Candidate \lya\ galaxies are those that appear in one of the narrow
band filters, but are undetected in the broad band filters. Given the
relatively low signal-to-noise ratios of the candidates, we also
require a detection in the intermediate band filter. This limits the
number of spurious candidates. We refer the interested reader to
Westra et al (2005, in prep.) for a description of the observations,
reduction and candidate selection.

\section{Confirmed \lya\ Emitter at $z=5.721$}
\label{sec:confirmedlya}

In a pilot study to test the effectiveness of the selection strategy
one of the brighter candidates (\mbox{\object{J114334.98-014433.9}},
hereafter \object{S11\_13368}) was observed with FORS2 on Yepun (UT4)
at the Cerro Paranal Observatory (Figure \ref{fig:thumbs}).

\begin{figure}[!htbp]
  \includegraphics[width=\columnwidth, trim=14pt 15pt 41pt 269pt, clip]{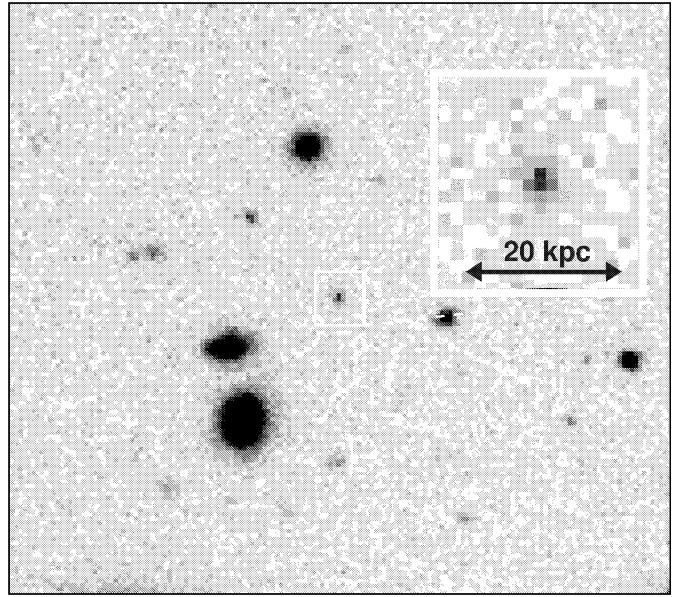}
\caption{
  A 30\arcsec\ x 30\arcsec\ region around the confirmed \lya\ galaxy
  \object{S11\_13368} at $z = 5.721$ from the pre-image taken with
  FORS2. The image has a pixel scale of 0\farcs252\,pix$^{-1}$.  The
  object is unresolved in this image. The seeing at the time of these
  observations was $\sim$\,0\farcs5. The exposure time was 3600
  seconds.}
\label{fig:pre}
\end{figure}

A pre-image with an intermediate-band filter (13\,nm) centered at
815\,nm was taken with FORS2 on 2004 February 16th in which
\object{S11\_13368} clearly was detected. Figure \ref{fig:pre} shows a
30\arcsec\ region around \object{S11\_13368}. The FWHM of stars in
this field are 0\farcs5 and \object{S11\_13368} is unresolved.

\begin{figure}[!htbp]
\centering
\includegraphics[trim=110pt 0pt 135pt 0pt,width=0.9\columnwidth,clip]{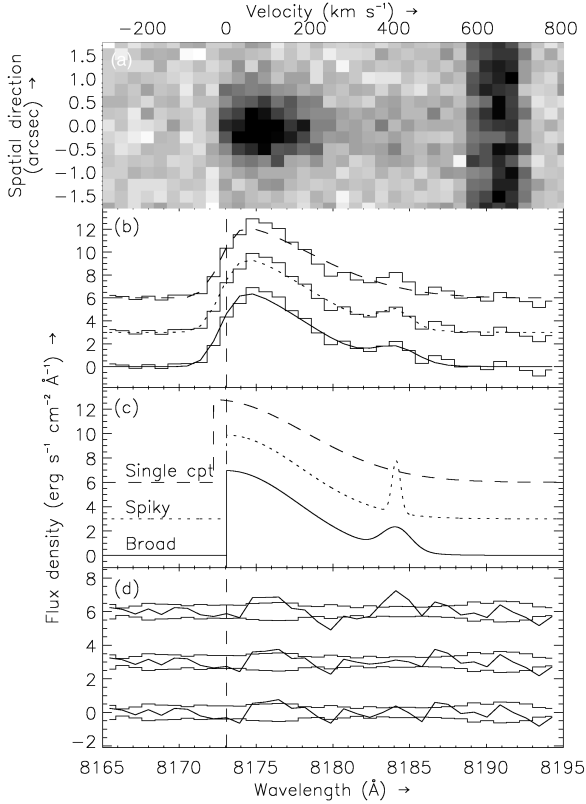}
\caption{
  A fit of both one- and two-component models to the profile of the
  \lya\ line. ($a$) The two dimensional spectrum centered on the \lya\ 
  line. In this unrectified spectrum, the night sky emission lines are
  not removed. The pixel scale for this image is
  0\farcs252\,pix$^{-1}$ in the spatial direction and
  0.86\,\AA\,pix$^{-1}$ in the dispersion direction. ($b$) Observed
  \lya\ line ({\it histograms}) with the three best-fitting models.
  For clarity, the models are offset by 3 \ergsPerAng. The
  two-component models consist of a two components: a broad, truncated
  Gaussian and a narrower redshifted Gaussian. The one-component model
  consists of only a broad, truncated Gaussian.  See Table
  \ref{tab:lineParameters} for the parameters of the `broad' ({\it
    solid}), `spiky' ({\it dotted}) and `single component' ({\it
    dashed}) models. ($c$) Same model line profiles as in ($b$) but
  before convolution with the instrument profile. ($d$) Observed data
  minus model fit (as plotted in ($b$)) residuals, demonstrating a
  random scatter about the zero flux line.  Also shown ({\it
    histograms}) is the 1$\sigma$-error spectrum from the observed
  data, which includes both sky- and Poisson noise. Note that the red
  peak is not \ion{N}{v}. At this redshift it would appear around
  8334\,\AA.}
\label{fig:1d}
\end{figure}

Three 1200\,sec exposures were taken on 2004 March 18th with FORS2
using the 1028z grism and a 1\arcsec\ slit. Frames were
bias-subtracted and flatfielded and were then combined with suitable
pixel rejection to remove cosmic rays. The 2D-spectrum (without
subtracting the sky lines) and the extracted sky-subtracted spectrum
are shown in Figure \ref{fig:1d}($a,b$), where one can clearly see a
single emission line with a broad red wing. No continuum is detected,
implying a 2$\sigma$ upper-limit for the continuum of
7$\times$10$^{-20}$\,\ergsPerAng\ over the rest frame wavelength range
1220 to 1230\,\AA.

\citet{Stern00} have reviewed the different ways in which high
redshift \lya\ can be verified and suggest that line asymmetry is the
surest way. Our spectral resolution \mbox{($R \sim 3600$)} is high
enough to securely confirm this asymmetry. This resolving power is
also more than adequate to rule out \Oii\ at $z=1.19$, since we do not
resolve the line into the close doublet. The separation of this
doublet at $z=1.19$ is 6.1\,\AA\ and therefore easily resolvable.
Similarly, we can rule out the possibility that the line is \Oiii\ at
$z=0.63$, due to the absence of the accompanying line in that doublet.
We can also rule out \ha\ at $z=0.25$. If the line was \ha, then we
should have either detected \NII\ and/or some flux in the $R$-band due
to the contributions of \hb, \Oiii\ and the continuum
\citep[e.g.][]{Kniazev04}. In Figure \ref{fig:thumbs}, one can see
that there is no detection in the $R$-band image. Given the absence of
all of these potential neighbouring features, and also the clear
asymmetry of the line, we identify it as \lya\ emission at $z=5.721$.

The integrated line flux of the line derived from the spectrum is
5$\times$10$^{-17}$~\lineunits, making it one of the brightest \lya\ 
emitting sources at this redshift
\citep[cf.][]{Ajiki03,Rhoads03,Maier03}. The 2$\sigma$ lower limit on
the rest frame equivalent width is $\sim$100\,\AA. At $z=5.721$, this
translates to a \lya\ luminosity of 1.8$\times$10$^{43}$~\ergsPersec\ 
suggesting an apparent star-formation rate of 16 \Msunyr, using the
conversion rate from \citet{Ajiki03}.

Following earlier works \citep{Dawson02,Hu04}, we fit both two- and
single-component models to the \lya\ line. The two-component fit
consists of a truncated Gaussian with complete absorption bluewards of
\lya\ line center and a redshifted Gaussian that is not truncated
\citep[e.g.][]{Hu04}. The one-component fit consists solely of a
truncated Gaussian (Figure \ref{fig:1d}{\it b,c}). Since the seeing
was narrower than the width of the slit, we convolve the model with a
Gaussian that has a FWHM of 2.3\,\AA. At 8175\,\AA\ this corresponds
to a resolution of $R\sim3600$. We use the Levenberg-Marquardt
nonlinear least-squares algorithm to find the best fit.
  
Two different two-component models fit the data with a similar reduced
$\chi_\nu^2$ values of $\sim$\,1.3. We refer to these two models as
the 'broad' model and the 'spiky' model. In the 'broad' model, the
redshifted component is broader and weaker and the central component
is narrower and stronger in comparison to the `spiky' model.  After
convolving with the instrumental profile both fits have similar
residuals. Both models have a central peak at a wavelength
corresponding to \lya\ at $z=5.721$. The redshifted component is
clearly detected in both models and lies $\sim$\,+400~\kms\ away from
the central peak. Given the similar reduced $\chi_\nu^2$, we cannot
favour one model over the other.
  
The single-component model has a broader main peak, which is slightly
bluer. This model does not fit the profile as well as the
two-component cases, particularly in the region of the red peak
($\chi_\nu^2$ = 2.2).

Table~\ref{tab:lineParameters} summarises the different model
components.

\begin{table}[!htbp]
\begin{center} 
\begin{tabular}{ccccccc}
\hline \hline\\
Component    & $\lambda_c$ & $f_{\rm peak}$ & \multicolumn{2}{c}{FWHM}  & \multicolumn{2}{c}{$\Delta v$}\\
(1)          & (2)         & (3)       & (4)           & (5)       & (6)    & (7)  \\ 
\hline\hline
\multicolumn{7}{c}{`Broad' model}\\
Main peak    & 8173.1      & 7.0    & 11.1          & 408       & \ldots & \ldots  \\
Red peak     & 8184.2      & 1.9    & 2.2           &  81       & +11.1  & 406 \\
\hline
\multicolumn{7}{c}{`Spiky' model}\\
Main peak    & 8173.1      & 6.9    & 11.4          & 419       & \ldots & \ldots  \\
Red peak     & 8184.1      & 4.3    & 0.6           &  24       & +11.0  & 405 \\
\hline
\multicolumn{7}{c}{`Single component' model}\\
Main peak    & 8172.2      & 6.8    & 14.0          & 514       & \ldots & \ldots  \\
\hline\hline\\
\end{tabular}
\caption{
  Model fit parameters as described in Section \ref{sec:confirmedlya}
  and indicated in Figure \ref{fig:1d}. Notes: (1) component of the
  fit, (2) central wavelength of the fitted component in \AA, (3) peak
  flux density in 10$^{-18}$~\ergsPerAng, (4) and (5) FWHM of full
  Gaussian of the profile in \AA\ and km\,s$^{-1}$, respectively, (6)
  and (7) line-of-sight outflow velocity in \AA\ and km\,s$^{-1}$,
  respectively.}
\label{tab:lineParameters}
\end{center} 
\end{table}

The integrated line fluxes for the two-component models are very
similar. If one were to include the flux that was missing from the
blue side of the truncated Gaussian, the total a line flux is
8.3$\times$10$^{-17}$~\ergsscm. This corresponds to a star-formation
rate of $\sim$\,27~\Msunyr, using the conversion rate from
\citet{Ajiki03}. The width on the blueward side of the profile is
solely due to instrumental broadening.

\section{Discussion}

We have presented a medium resolution spectrum of a bright \lya\ 
emitting galaxy at $z=5.721$. The spectrum consists of a single
emission line and no continuum. The line shows a distinct asymmetry,
which undoubtedly confirms it as \lya.  We model the line with two
components: a one-sided Gaussian and a narrower, redshifted component.
The profile of the blue side of the line is entirely defined by the
instrument profile.

Generally, the second component is less frequently observed, although
it is possible that it has been missed in the spectra of other \lya\ 
emitting galaxies. Most of these spectra were taken at lower
resolution and are considerably noisier. A second peak in the \lya\ 
line is a clear signature of an expanding shell of neutral hydrogen
\citep{Dawson02,Ahn03}. The strength and shape of the secondary peak
depends on the kinematics and the quantity of neutral hydrogen in the
expanding shell and the amount and distribution of dust throughout the
galaxy \citep{Ahn04}.

\object{S11\_13368} appears to be a brighter and more distant example
of \object{J1235.8+6215} at $z=5.190$ \citep{Dawson02}. The line
profile is strongly asymmetric in both objects, and both suggest a
second redshifted component. Both objects are also very compact.
However, there are some noteworthy differences.  The \lya\ line in
\object{S11\_13368} is considerably broader, and the redshifted
component is a lot narrower, even in our `broad' model.

The intrinsic \lya\ profile is heavily modified by the surrounding gas
and the fraction of the line that is finally observed is very model
dependent \citep{Santos04a,Ahn04}. In general, it is only a fraction
of the intrinsic flux. Hence, star formation rates that are estimated
from the observed \lya\ flux directly, as they are done in this paper,
could drastically underestimate the true star formation rate.
Similarly, the centroid of the observed profile is also model
dependent. This directly leads to an uncertainty in the redshift of
about 0.01, if no other lines are visible, which is usually the case
for such high redshift galaxies.

\object{S11\_13368}, like \object{J1235.8+6215}, is very compact. With
a projected size of $\sim3$\,kpc or less, it is comparable to the size
of the star forming regions in local starbursting galaxies; however,
the star formation rate is much higher. Not all \lya\ emitting
galaxies at $z\sim\,5.7$ are as compact. In Figure 5, we plot apparent
size of the star forming region versus the inferred star formation
rate for a sample of local starbursts and distant galaxies.  The
emission line region in \object{LAE J1044-0130} occurs over a region
that is an order of magnitude larger than emission line regions in
\object{S11\_13368} and \object{J1235.8+6215} even though the inferred
star formation rate is significantly less. Given that the projected
star formation rate per unit area in \object{S11\_13368} far exceeds
0.1 M$_\odot$\,yr$^{-1}$\,kpc$^{-2}$, it is likely that a hot,
enriched starburst-driven gas is outflowing into the halo of
\object{S11\_13368}, facilitating the enrichment of the halo and the
escape of Lyman continuum photons \citep{Heckman00,TenorioTagle99}.

\begin{figure}[!htbp]
\centering
\includegraphics[width=0.90\columnwidth,trim=90pt 0pt 120pt 15pt,clip]{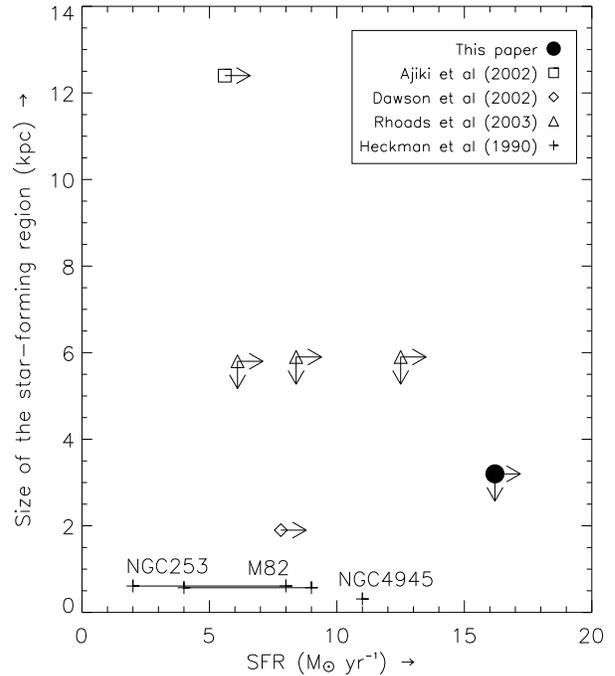}
\caption{
  Size of the star-forming region versus the star-formation rate. The
  SFR of two galaxies from \citet{Heckman90} is indicated by a bar, as
  there are two measurements of the SFR. The arrows represent
  lower-limits to the SFR and upper-limits to the size of the major
  axis, (due to seeing limited observations).  Error-bars are not
  included. The plus signs represent three well-known local
  starbursting galaxies \citep{Heckman90}.}
\label{fig:sfr}
\end{figure}

\begin{acknowledgements}
  The authors wish to thank the Max-Planck-Institut f\"ur Astronomie
  and the DDT grant of the European Southern Observatory for providing
  the narrow band filters which are crucial to the WFILAS survey. We
  also like to thank the anonymous referee for his/her useful
  suggestions and comments, which made us improve the article a lot.
  D.~H. Jones is supported as a Research Associate by Australian
  Research Council Discovery-Projects Grant (DP-0208876), administered
  by the Australian National University. C. Wolf was supported by a
  PPARC Advanced Fellowship. Reduction was done with IRAF, which is
  distributed by the National Optical Astronomy Observatories, which
  are operated by the Association of Universities for Research in
  Astronomy, Inc., under cooperative agreement with the National
  Science Foundation.
\end{acknowledgements}

\bibliographystyle{aa.bst} \bibliography{wfilas.bib,astroph.bib}

\end{document}